\documentclass{aa}
\usepackage{epsf,graphicx}

\begin{document}

  \thesaurus{06.          
              (08.14.1;   
               02.04.1)   
            }

\title{ Cooling neutron stars with localized protons }
\author{D.A. Baiko\inst{1} \and P. Haensel\inst{2}}
\offprints{P. Haensel}

\institute{
         A.F.\ Ioffe Physical Technical Institute,
         Politekhnicheskaya 26, 194021, St.Petersburg, Russia \\
	 e-mail: baiko@astro.ioffe.rssi.ru
	 \and
	 N.\ Copernicus Astronomical Center,
         Bartycka 18,  00-716, Warsaw, Poland \\
         e-mail: haensel@camk.edu.pl
	  }
\date{Accepted 20th January 2000 }
\maketitle



\begin{abstract}
We analyze cooling of neutron stars,
assuming the presence of localized  protons in the densest
region of their cores.  Choosing a single threshold
density for proton localization and adjusting neutron star
mass, we reproduce the observational
data on effective surface temperatures of Vela and
PSR 0656+14, with or without an accreted
hydrogen envelope. However,
the presence of a tiny hydrogen envelope is mandatory, in this model,
for reproducing the Geminga data.

\end{abstract}

The model of dense neutron star matter with localized protons
was proposed by Kutschera \& W{\'o}jcik (\cite{KW93}).
According to these authors, in high-density matter 
with low proton fraction 
the ``zero-point'' (Fermi) kinetic energy of protons, estimated
from the uncertainty principle,
becomes small, compared to the energy of proton interactions
with density waves 
of the neutron background. 
These  density waves
resemble hydrodynamic sound or long wavelength
acoustic phonons in a solid.  
The coupling of protons to the neutron density waves  
results in a significant increase of proton effective mass
in close analogy with polaron behaviour of a slow electron
in a polar solid which also acquires a large effective mass because
of the interactions with lattice phonons. 

Moreover, considering one proton
in neutron matter, Kutschera \& W{\'o}jcik have shown
that above some critical density
$\rho_{\rm lp} \ga 4 \rho_0$
($\rho_0 = 2.8 \times 10^{14}$ g cm$^{-3}$)
the energy of a quantum state, in which a proton wave function
is localized around some spatial point associated with
the neutron density minimum, is lower than the energy of a
state, where proton wave function is nonlocalized
and the distribution of neutrons is uniform.
Such a state also has an analog 
in  solid state physics, 
the so-called small polaron state.
It occurs when an electron interacts with 
a lattice so strongly that the latter deforms and this 
deformation traps the electron in a self-consistent manner.

This result by Kutschera \& W{\'o}jcik
may be applied to neutron star cores,
where one has finite admixture of protons (rather than one proton
in neutron matter), if the proton fraction $x_p$ is so small that
wave functions of neighbouring localized protons do not overlap.
Small proton fractions in high density neutron-star matter
were obtained by Wiringa et al.\ (\cite{WFF88}), who performed extensive
many-body calculations of the ground state, based on
variational method and assuming realistic nuclear hamiltonian.
In particular, their UV14+TNI
model, which will be used further, predicts
a proton fraction below 5\% for densities $\rho > 3\rho_0$,
and, eventually, complete disappearance of protons at
about $7 \rho_0$. 
This result is in contradiction
with some other calculations of equation of state (EOS) of neutron
star matter [based on the relativistic mean-field approach,
see Glendenning (\cite{G96}), or performed within the
Brueckner-Bethe-Goldstone many-body theory, 
e.g., Baldo et al.\ (\cite{BBB97})] which
yield a proton fraction growing monotonically with
baryon density $n_b$. 
The high density behaviour of $x_p(n_b)$ is,
therefore, subject to a significant theoretical uncertainty.
Arguments  in favor of $x_p(n_b)$, which decreases and
eventually vanishes at high $n_b$, were presented by
Kutschera (\cite{K94}).

The localization of protons (if confirmed)
leads to dramatic changes of basic kinetic properties
of matter in the neutron star core. The reason is that
the localized protons are very efficient
scatterers of the main transport agents, electrons and neutrons.
In effect, in the low-temperature regime, 
the thermal conductivity $\kappa$, for instance,
behaves as $\propto T$, instead of the conventional
$T^{-1}$ dependence when scattering is provided by degenerate
particles. As shown by Baiko \& Haensel (\cite{BH99})
(hereafter Paper I),
this extends the time scale
of thermal equilibration in the neutron star core by
about two orders of magnitude independently of the state
of neutrons (normal or superfluid).

The neutrino emissivity of matter with localized protons is
also very non-standard. First of all,
the conventional beta-processes which involve charged
currents (direct or modified Urca processes) are forbidden
since at given baryon density the minimum of energy
corresponds to a specific proton fraction, while any process which
changes this fraction would require rearrangement
of the whole state and, consequently,
much larger energy than the thermal energy available, say,
at $T \la 10^9$ K. Thus, the only possible processes are
those which involve neutral currents. These are
bremsstrahlung of neutrino--antineutrino pairs in $nn$,
$np$, and $ep$ collisions.

The kinetic properties of matter with localized protons
were considered in Paper I. In particular, 
simple analytical formulae were derived for the electron
and neutron thermal conductivities and shear viscosities,
for the electron electrical conductivity as well as
for the neutrino emissivity
from $np$ and $ep$ bremsstrahlung. All these quantities were
calculated under two simplifying assumptions. Firstly,
it was assumed that the localized protons did not exibit
any correlations
(i.e., behaved 
like impurities).
Secondly, the in-vacuum matrix elements of strong interactions
were used. Both assumptions
lead to an increase of the reaction rates by a factor of 
a few,  
since it is
generally expected that the medium effects
and correlation of scatterers
reduce the matrix
elements of elementary scattering processes,
at least for small momentum transfers.
Thus the results of Paper I give
the upper bounds to the neutrino emissivities and the lower
bounds to the transport coefficients. In other words, they represent the
maximum effect that one can expect from the proton localization.

The aim of the present paper is to apply the results of Paper I
to modelling neutron star cooling in order to check if
the hypothetical proton localization is compatible with
observational data on the neutron star effective surface
temperatures. To do that we have to specify
the neutron star model.
In the core
(at densities $\rho > \rho_{\rm cr} = 2.5 \times 10^{14}$ g cm$^{-3}$),
we take the UV14+TNI EOS by Wiringa et al.\ (\cite{WFF88}),
while in the crust (below $\rho_{\rm cr}$) we take the EOS by 
Negele \& Vautherin
(\cite{NW73}) 
above neutron-drip density, 
and the EOS by Baym et al. (\cite{BPS71}) below neutron-drip density.  
The model is then
determined by the only parameter, the central mass density $\rho_{\rm c}$,
and is
constructed by integrating the Tolman-Oppenheimer-Volkoff equation
outward the center.

We use the cooling code 
described, e.g., in
Levenfish et al.\ (\cite{LSY99}).
The code includes explicitly effects of General Relativity (GR)
and traces a steady thermal evolution of
a spherically symmetric neutron star. Matter at densities
$\rho > \rho_{\rm b} = 10^{10}$ g cm$^{-3}$ is assumed to be
isothermal with constant temperature $T^{\infty}_{\rm i}$
(as measured at infinity). The local
temperature, $T_{\rm i}$, appears to be dependent
on the radial coordinate $r$ due to the GR effect.
The cooling is governed by the thermal balance equation with energy losses
due to neutrino emission
from the core and photon emission from the stellar surface.
The local effective surface temperature $T_{\rm s}$
is obtained from the local internal
temperature, $T_{\rm b}$, at $\rho = \rho_{\rm b}$ using the formula
by Potekhin et al.\ (\cite{PCY97}). The latter formula fits
the results of numerical simulations of heat transport through the crust.
The formula depends also on the mass $\Delta M$ of the outer envelope 
made of light elements (light envelope). 
Such an envelope may, in principle,  reside on the neutron 
star surface and due to its higher thermal conductivity 
the envelope would decrease the thermal insulation of the inner stellar 
regions and modify the $T_{\rm s}(T_{\rm b})$ relationship. 
The surface temperature $T^{\infty}_{\rm s}$,
as measured by a distant observer, is related to $T_{\rm s}$ as
$T_{\rm s}^{\infty} = T_{\rm s} \, \sqrt{1-2GM/Rc^2}$,
where $M$ and $R$ are, respectively,
the mass and radius of the neutron star.

\begin{figure}[t]
\begin{center}
\leavevmode
\includegraphics[height=85mm,bb= -1 11 346 346]{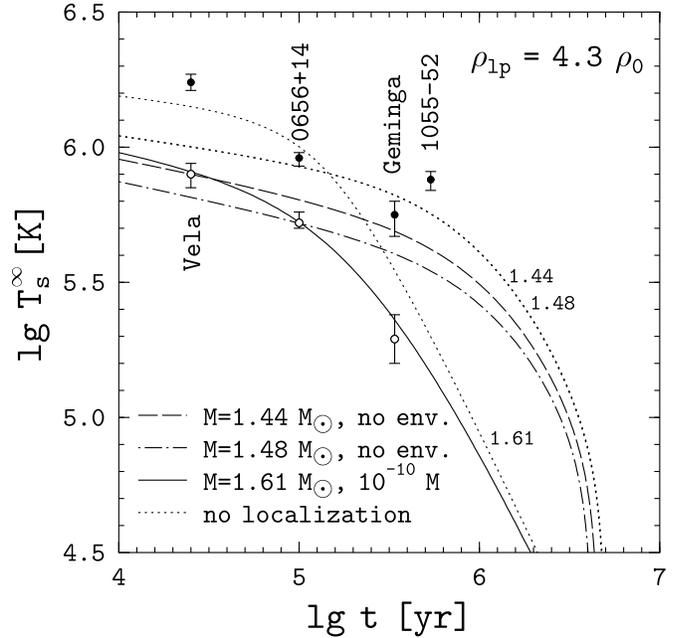}
\end{center}
\vspace{-0.4cm}
\caption[]{Redshifted surface temperature versus 
neutron star age. Filled circles represent interpretation
of observations with the black-body model while open
circles are the ``atmospheric'' interpretations.
The threshold parameters for proton localization
are $x_{p0}=5\%$ and $\rho_{\rm lp}=4.3 \rho_0$.
The solid curve corresponds to the stellar model with an
envelope of light elements, while the dashed and dot-dashed 
curves are calculated assuming no light envelope.
The dotted curves are for the same star models (masses, presence or absence
of the envelope) but without proton localization.
}
\label{cool}
\end{figure}

For our purposes we have introduced a number of modifications
into the cooling code. All the changes are related to matter
at densities above $\rho_{\rm lp}$. First of all, we have removed
the proton contribution to the heat capacity. Then
we have revised contributions to neutrino emissivity.
Above $\rho_{\rm lp}$, if neutrons are nonsuperfluid
(which is the case in study), the main contribution to
the neutrino emissivity is due to $nn$ and $np$ neutrino-pair
bremsstrahlung. The rate of the $nn$ process is not changed by the
proton localization, while the emissivity of the $np$ process
reads (Paper I):
\begin{eqnarray}
   Q^{np}_{\rm Brem}
   &=& 1.3 \times 10^{21} \,
            \left( {n_b \over 4 n_0} \times {x_p \over 0.01} \right) \,
            \left( {n_n \over 4 n_0} \right)^{4/3}
            \left( m^\ast_n \over m_n \right)^2 \,
\nonumber \\
      &\times& S^{\rm t}_{np} \, T^6_9
            \quad {\rm ergs \; s^{-1} \; cm^{-3}},
\label{Qnnum}
\end{eqnarray}
where $x_p$ is the fraction of protons by
number, $n_n$ is the neutron number density, $m_n$ and $m_n^\ast$
are the bare and effective neutron masses, respectively;
$n_0 = 0.16$ fm$^{-3}$,
and $T_9$ is (local) temperature in units of $10^9$ K. 
Finally, $S^{\rm t}_{np}$
is a function of $n_n$ related to the total $np$ cross-section
in vacuum,
\begin{equation}
            S^{\rm t}_{np} =
            26.09 + 3.444 \, {4 n_0 \over n_n}.
\label{U2fit}
\end{equation}
%

\begin{figure}[t]
\begin{center}
 \leavevmode
\includegraphics[height=85mm,bb= -1 11 346 346,clip]{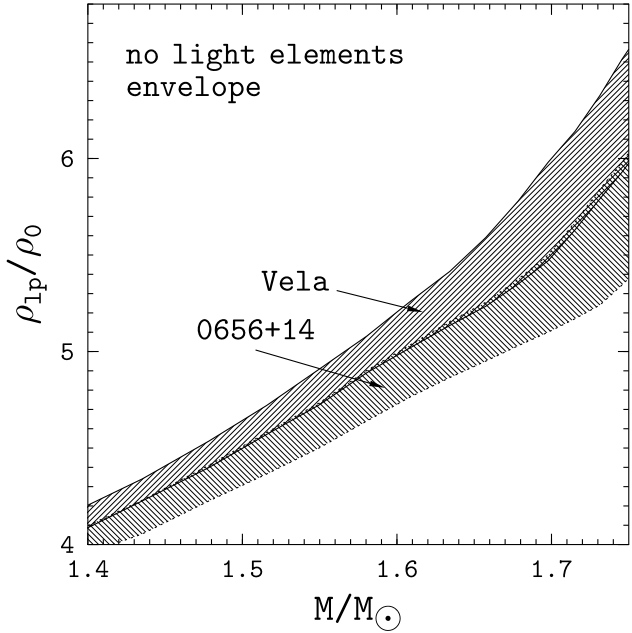}
\end{center}
\vspace{-0.4cm}
\caption[]{Threshold density of localized protons versus mass of a neutron
star for cooling model without envelope of light elements on the
surface. The dotted and solid lines confine the 
parameter domains corresponding
to the $T_{\rm s}^\infty$ error bars of  PSR 0656+14 and
Vela, respectively.
}
\label{noenv}
\end{figure}

Typical cooling curves obtained within this model are presented in 
Fig.\ \ref{cool}. The effect of the proton localization can be 
easily seen 
by comparing the dotted curves,  
which represent
the cooling without the proton localization, with curves of other types, 
which show a faster cooling due to the localization effect.
The rate 
 of cooling without the proton localization is virtually independent
of the mass: the dotted curves for 1.44 $M_\odot$ and 1.48 $M_\odot$
stars coincide. If the effect of the proton localization is taken into account, 
the more massive star cools more rapidly because at fixed threshold density,
$\rho_{\rm lp}$, the more efficient neutrino pair bremsstrahlung in 
neutron-localized proton collisions is operative in a larger region of the
neutron star core leading to a larger overall neutrino luminosity.

\renewcommand{\arraystretch}{1.4}

The results of our cooling simulations may be used for interpretation
of thermal emission from the surface of cooling neutron stars
of ages $t \ga 10^4$ yr,
which is the typical thermal equilibration time for 
neutron stars with
localized protons (Paper I). 
\begin{table}
\begin{eqnarray}
\begin{array}{|c|c|c|c|c|}
\hline
    & {\rm Vela} & 0656{\rm +}14 & {\rm Geminga} & 1055{\rm -}52 \\
\hline
  t, 10^4~{\rm yrs} &  2.5^{a)}  &11 & 34 & 54 \\
\hline
  T^\infty_{\rm bb}, 10^5~{\rm K}& 17.4 {\rm \pm} 1.2 & 9.1 {\rm \pm} 0.5 &  
  5.6^{+0.7}_{-0.9} & 7.5 {\rm \pm} 0.6 \\
\hline
  T^\infty_{\rm bb}, {\rm Ref} & [{\rm OFZ}93] & [{\rm PMC}96] & 
  [{\rm HW}97] & [{\rm OF}93] \\
\hline
  T^\infty_H, 10^5~{\rm K} & 7{\rm -}8.6 & 5.3^{+0.4}_{-0.3} & 2{\rm -}3 &
 - \\
\hline  
  T^\infty_H, {\rm Ref} & [{\rm PSZ}96] & [{\rm ACPRT}93] & [{\rm MPM}94] & 
 - \\
\hline
\end{array}
\nonumber
\end{eqnarray}
\caption[]{The observational data on the 4 middle-aged 
isolated neutron stars.

  $^{a)}$  according to Lyne et al.\ (\cite{Lyne96});

[{\rm OFZ}93] {\"O}gelman et al.\ (\cite{OFZ93});

[{\rm PMC}96] Possenti et al.\ (\cite{PMC96}), 90\% confidence level (c.l.);

[{\rm HW}97] Halpern \& Wang (\cite{HW97}), 90\% c.l.;

[{\rm OF}93] {\"O}gelman \& Finley (\cite{OF93});

[{\rm PSZ}96]  Page et al.\ (\cite{PSZ96}), 90\% c.l., for adopted lower
limit of distance $d=300$ pc;

[{\rm ACPRT}93] Anderson et al.\ (\cite{Aal93});

[{\rm MPM}94] Meyer et al.\ (\cite{MPM94}), typical value of the surface
redshift factor, $T_{\rm s}^\infty/T_{\rm s}=0.8$, were used.
}
\label{tabl}
\end{table}
At present there are 4 sources, identified as isolated
neutron stars,
which show thermal soft X-ray emission and have
characteristic ages $t$ in excess of $10^4$ yrs. 
These are Vela, PSR 0656+14, Geminga and PSR 1055-52.
The data that we use are summarized in Table \ref{tabl}.
The effective surface temperatures $T^\infty_{\rm s}$ are determined from 
fitting the observed spectra of the sources in two different ways,
either by black-body spectra ($T^\infty_{\rm bb}$ in Table \ref{tabl}) 
or by spectra obtained in realistic 
models of magnetized neutron star atmospheres 
($T^\infty_H$ in Table \ref{tabl}
for magnetized hydrogen atmosphere).
The inferred values of $T^\infty_{\rm s}$ appear to be quite different.
Radiation emerging from
an atmosphere composed of light elements
(hydrogen and helium) has harder spectrum than the black-body
radiation with the same $T^\infty_{\rm s}$ due to the
decrease of opacity at higher energies (Romani \cite{R87}).
Consequently, the effective
temperatures predicted by the atmosphere models are
2-3 times lower than the black-body ones
which require lower $d/R$ (distance
to radius) ratio. With the stellar radius fixed
around the standard value of 10 km this translates into about
10 times smaller distances to the objects. The atmospheres
composed of iron, on the contrary, produce radiation with
a softer spectrum which is more similar on average to
that of a black body.

The magnetic fields inferred from
the spin-down rates are above $10^{12}$ G for all 4 stars.
Such fields must have significant effect on
the atmospheric opacities and therefore will modify 
the emerging spectrum (Shibanov et al.\ \cite{SZPV92}).
In general, the magnetic field makes the spectrum softer
and somewhat elevates the effective temperature compared to the
non-magnetized atmosphere case. However,
no accurate magnetized atmosphere model have been developed so far
for the relevant temperature range ($T_{\rm s} < 10^6$ K)
where one should take into account effects of
atom motion on the opacity
(e.g., Pavlov \& Zavlin, \cite{PZ98}). These effects were neglected
while obtaining the values reproduced in Table \ref{tabl}.

The surface temperatures $T^\infty_{\rm s}$ are plotted in
Fig.\ \ref{cool}. The filled and open circles correspond
to black-body models and simplified magnetized hydrogen 
atmosphere models, respectively. 
As one concludes from Table \ref{tabl} and the figure, the 
overall theoretical 
uncertainty in surface temperatures appears to be quite large;
for instance $T^\infty_{\rm s} \sim (2{\rm -}6) \times 10^{5}$ K for Geminga. 
This does not allow one to draw any definite
conclusion about the cooling scenario of the neutron star
except that it probably rejects the ``rapid''
cooling (via direct Urca process unsuppressed by
superfluidity in dense matter). We note that
the surface temperatures obtained from the hydrogen
atmosphere models are rather low and cannot be explained
within the ``standard'' cooling scenario
(the neutrino losses via modified Urca process
from non-superfluid matter). 
On the other hand, such a scenario can explain
the black-body temperatures.
Thus, the ``atmospheric'' temperatures (if the true temperatures
turn out to be close to them) may provide
more stringent test of the theory of neutron-star interior.
For this reason let us adopt
the ``atmospheric'' interpretation of observations and
focus on lower error bars of Vela, PSR 0656+14,
and Geminga.

\begin{figure}[t]
\begin{center}
\leavevmode
\includegraphics[height=85mm,bb=-1 10 346 346,clip]{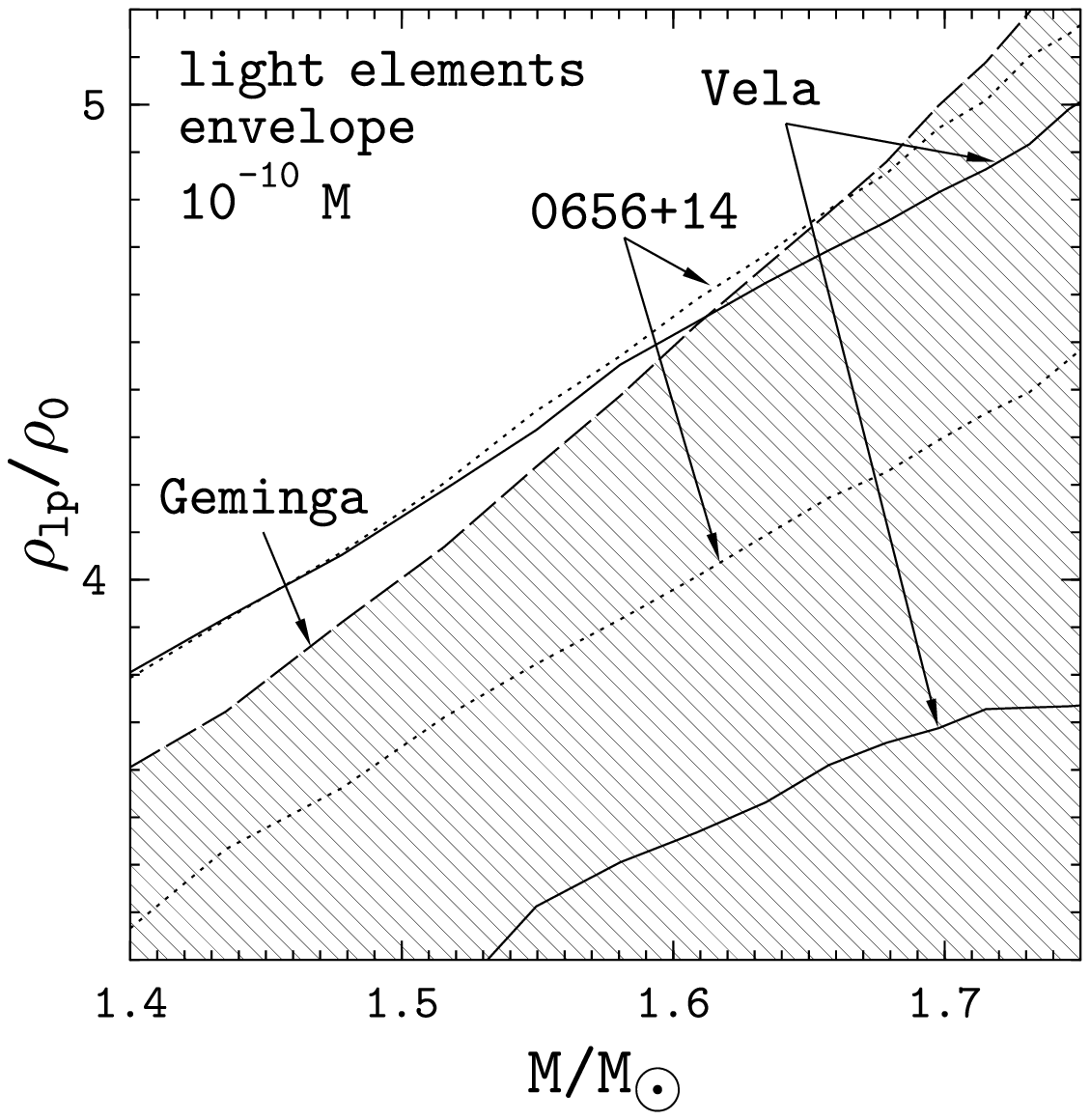}
\end{center}
\vspace{-0.4cm}
\caption[]{Same as in Fig.\ \ref{noenv} but for the model of a neutron
star with envelope of light elements (of mass
$10^{-10}\, M$) on the surface.
The dashed line confines the (shaded) parameter domain allowed for Geminga.
}
\label{env}
\end{figure}

Having fixed EOS we are left with three parameters which influence
the cooling. These are the central stellar density $\rho_{\rm c}$ and the
threshold values of mass density, $\rho_{\rm lp}$,  and proton fraction,
$x_{p0}$. In addition, we can vary $\Delta M$, the mass 
of the surface envelope made of light elements. 
Note, that the presence of
the hydrogen atmosphere is not equivalent
to the presence of a light envelope
as the amount of hydrogen needed to modify
the emerging spectrum 
(1-10 g cm$^{-2}$ or $\sim 10^{-20}{\rm -}10^{-19} M_\odot$)
is much lower
than that required to
change the $T_{\rm s}(T_{\rm b})$ relationship.
Let us fix
the threshold proton fraction $x_{p0}= 5\%$,
and consider two models: without envelope, Fig.\ \ref{noenv},
and with a light envelope of mass $\Delta M = 10^{-10}M$,
Fig.\ \ref{env}. The strips bounded by the lines of various types 
in Figs.\ \ref{noenv} and
\ref{env} correspond to the domain of $\rho_{\rm lp}$
and $M$, for which cooling curves cross the error bar of a given source.
Let us stress that Geminga (whose parameter domain is shaded in
Fig.\ \ref{env}) can be explained 
only assuming 
the light envelope. In the latter case and for realistic $\rho_{\rm lp}$
(presumably above $4 \rho_0$) the mass of Geminga should be above
1.5 $M_{\odot}$. The other two sources can be explained
either with or without the envelope but for fixed $\rho_{\rm lp}$
their masses in the model with envelope should be
higher. Finally let us mention that the strips are not very sensitive
to the envelope mass for $\Delta M \ga 10^{-10}M$.

Fig.\ \ref{cool} illustrates the cooling curves for the fixed
threshold density $\rho_{\rm lp} = 4.3 \rho_0$. The solid curve
represents the model with the light envelope and $M=1.61 M_\odot$
($\rho_{\rm c} = 5.4 \rho_0$). It crosses the ``atmospheric'' 
error bars of Vela,
PSR 0656+14 and Geminga simultaneously. 
The dashed and dot-dashed curves are
envelope-free models. The masses should be lower, for instance,
$1.44$ and $1.48 M_{\odot}$ for Vela and PSR 0656+14, respectively.

We have performed cooling simulations of neutron stars,
with realistic equation of state, assuming localization of
protons  above  some threshold density.
These results have been used for interpretation of 
effective surface temperatures of observed isolated, middle-aged
neutron stars  ($t \geq 10^4$ yr).
We have shown that the available observational data are consistent
with the proton localization and can be reproduced if one chooses 
model parameters within the specific domains in the $\rho_{\rm lp}-M$ 
plane. Choosing a single threshold
density for proton localization and adjusting neutron star
mass, we reproduce the observational
data for Vela and  PSR 0656+14, with or without an
accreted, hydrogen envelope. However,
the presence of the hydrogen envelope is required in this model 
for explaining the observations of Geminga.

In conclusion, we note that
the localization of protons is not necessary for 
the explanation 
of the
available data on neutron star effective surface temperatures.
Nevertheless, as is seen from the Fig.\ \ref{cool}, 
it changes significantly 
the cooling rate 
and, if its existence is confirmed, 
it represents an important effect 
to be taken into account in realistic
neutron star cooling 
calculations.

\acknowledgements
We are grateful to D.G.\ Yakovlev and Yu.A.\ Shibanov for enlightening
discussions. Special thanks go to K.P.\ Levenfish and O.Yu.\ Gnedin
for assistance with the cooling code. The work was supported in part by 
RFBR (grant 99-02-18099), INTAS (96-0542), and KBN (2 P03D 014 13).

\end{document}